\documentclass[twocolumn,english,prl,twocolumn,floatfix,footinbib,superscriptaddress,showpacs, showkeys,notitlepage]{revtex4-1}
\usepackage[latin9]{inputenc}
\usepackage{amsmath}
\usepackage{graphicx}
\usepackage{esint}

\makeatletter

\@ifundefined{textcolor}{}
{%
 \definecolor{BLACK}{gray}{0}
 \definecolor{WHITE}{gray}{1}
 \definecolor{RED}{rgb}{1,0,0}
 \definecolor{GREEN}{rgb}{0,1,0}
 \definecolor{BLUE}{rgb}{0,0,1}
 \definecolor{CYAN}{cmyk}{1,0,0,0}
 \definecolor{MAGENTA}{cmyk}{0,1,0,0}
 \definecolor{YELLOW}{cmyk}{0,0,1,0}
}

\global\long\def\ket#1{\left| #1\right\rangle }

\global\long\def\bra#1{\left\langle #1 \right|}

\global\long\def\kket#1{\left\Vert #1\right\rangle }

\global\long\def\bbra#1{\left\langle #1\right\Vert }

\global\long\def\braket#1#2{\left\langle #1\right. \left| #2 \right\rangle }

\global\long\def\bbrakket#1#2{\left\langle #1\right. \left\Vert #2\right\rangle }

\global\long\def\av#1{\left\langle #1 \right\rangle }

\global\long\def\tr{\text{Tr}}

\global\long\def\im{\text{Im}}

\global\long\def\re{\text{Re}}

\newcommand{\be}{\begin{eqnarray}}\newcommand{\ee}{\end{eqnarray}}\def\beq{\begin{equation}}\def\eeq{\end{equation}}

\usepackage{babel}

\usepackage{babel}

\makeatother

\usepackage{babel}
\begin{document}

\title{Non-Markovian effects in electronic and spin transport}

\author{Pedro Ribeiro}

\affiliation{Russian Quantum Center, Novaya street 100 A, Skolkovo, Moscow area,
143025 Russia}

\affiliation{Centro de Física das Interações Fundamentais, Instituto Superior
Técnico, Universidade de Lisboa, Av. Rovisco Pais, 1049-001 Lisboa,
Portugal}

\author{Vitor R. Vieira}

\affiliation{Centro de Física das Interações Fundamentais, Instituto Superior
Técnico, Universidade de Lisboa, Av. Rovisco Pais, 1049-001 Lisboa,
Portugal}
\begin{abstract}
We derive a non-Markovian master equation for the evolution of a class
of open quantum systems consisting of quadratic fermionic models coupled
to wide-band reservoirs. This is done by providing an explicit correspondence
between master equations and non-equilibrium Green's functions approaches.
Our findings permit to study non-Markovian regimes characterized by
negative decoherence rates. We study the real-time dynamics and the
steady-state solution of two illustrative models: a tight-binding
and an XY-spin chains. The rich set of phases encountered for the
non-equilibrium XY model extends previous studies to the non-Markovian
regime. 
\end{abstract}


\maketitle
\begin{minipage}[t]{1\columnwidth}%
\global\long\def\ket#1{\left| #1\right\rangle }

\global\long\def\bra#1{\left\langle #1 \right|}

\global\long\def\kket#1{\left\Vert #1\right\rangle }

\global\long\def\bbra#1{\left\langle #1\right\Vert }

\global\long\def\braket#1#2{\left\langle #1\right. \left| #2 \right\rangle }

\global\long\def\bbrakket#1#2{\left\langle #1\right. \left\Vert #2\right\rangle }

\global\long\def\av#1{\left\langle #1 \right\rangle }

\global\long\def\tr{\text{tr}}

\global\long\def\Tr{\text{Tr}}

\global\long\def\pd{\partial}

\global\long\def\im{\text{Im}}

\global\long\def\re{\text{Re}}

\global\long\def\sgn{\text{sgn}}

\global\long\def\Det{\text{Det}}

\global\long\def\abs#1{\left|#1\right|}

\global\long\def\up{\uparrow}

\global\long\def\down{\downarrow}

\global\long\def\k{\mathbf{k}}

\global\long\def\wks{\mathbf{\omega k}\sigma}

\global\long\def\vc#1{\mathbf{#1}}

\global\long\def\bs#1{\boldsymbol{#1}}
\end{minipage}Out-of-equilibrium open quantum systems in contact with thermal reservoirs
are fundamentally different from isolated autonomous systems. Thermodynamic
gradients, such as temperature and chemical potential differences,
may induce a finite flow of particles, energy or spin, otherwise conserved
quantities. 

The interest in out-of-equilibrium processes has been boosted in recent
years by considerable experimental progress in the manipulation and
control of quantum systems under non-equilibrium conditions in as
cold gases \cite{Kinoshita2006,Bloch2008}, nano-devices \cite{Bonilla2005,Dubois2013}
 and spin \cite{Zutic2004,Khajetoorians2013} electronic setups. This
renewed attention in non-equilibrium processes has raised a number
of new questions, such as the existence of intrinsic out-of-equilibrium
phases and phase transitions \cite{Mitra2006,Prosen2008,DallaTorre2010,Kirton2013,Prosen2011},
the definition of effective notions of temperature \cite{Hohenberg1989,Cugliandolo1997,Sonner2012,Ribeiro2013b,Torre2013},
universality of dynamics after quenches \cite{Calabrese2006,Karrasch2012,Shchadilova2014}
and thermalization \cite{Deutsch1991,Srednicki1994,Rigol2008}.

Among the set of theoretical tools available to tackle non-equilibrium
quantum dynamics \cite{Eckstein2010,Arrigoni2013}, the Kadanoff-Baym-Keldysh
non-equilibrium Green's functions formalism allows for a systematic
derivation of the evolution from the microscopic Hamiltonian of the
system and its environment. An alternative approach consists on treating
open quantum systems with the help of master equations for the reduced
density matrix $\rho$. The formalism is generic as any process describing
the evolution of a system and its environment can be effectively described
by a master equation of the form \cite{H.-P.BreuerandF.Petruccione2002}
\begin{multline}
\pd_{t}\rho=\mathcal{L}_{t}\rho=-i\left[H\left(t\right),\rho\right]+\\
\sum_{\ell}\gamma_{\ell}\left(t\right)\left[L_{\ell}\left(t\right)\rho L_{\ell}^{\dagger}\left(t\right)-\frac{1}{2}\left\{ L_{\ell}^{\dagger}\left(t\right)L_{\ell}\left(t\right),\rho\right\} \right]\label{eq:Lindblad}
\end{multline}
where the $L_{\ell}$'s are a suitable set of jump operators, which,
without loss of generality, satisfy $\tr\left[L_{\ell}\left(t\right)\right]=0$
and $\tr\left[L_{\ell'}^{\dagger}\left(t\right)L_{\ell}\left(t\right)\right]=\delta_{\ell\ell'}$,
and $H$ is the system's Hamiltonian \cite{Hall2014}. The specific
form of the $L_{\ell}$'s is only known for rather specific examples
\cite{Kanokov2005,Vacchini2010,Rivas2010a}. To use this approach
on a practical level one has to rely on various approximations that
restrict its application range \cite{Gurvitz1996,Gurvitz1998}. Trace
preservation, which Eq.(\ref{eq:Lindblad}) respects, and positivity
are essential in order for $\rho\left(t\right)$ to represent a physically
allowed density matrix. Generic conditions on $L_{\ell'}\left(t\right)$
and $\gamma_{\ell}\left(t\right)$ to ensure that the complete positivity
of $\rho\left(t\right)$ is maintained throughout the evolution are
yet unknown \cite{Rivas2010a}. For the case where all decoherence
rates are non-negative ($\gamma_{\ell}\left(t\right)\geq0$) positivity
can be proven \cite{Gorini1976a,Lindblad1976}. This condition implies
that the operator $\mathcal{E}_{t,t'}\left(\rho\right)=Te^{\int_{t'}^{t}d\tau\mathcal{L}_{\tau}}\rho$
(where $T$ stands for the time-ordered product) is a completely positive
map for all $t>t'>0$. In this case $\mathcal{E}_{t,t'}$ is also
contractive, i.e. $\pd_{t}D\left[\mathcal{E}_{t,t'}\left(\rho_{1}\right),\mathcal{E}_{t,t'}\left(\rho_{2}\right)\right]\leq0$,
for a suitable measure of distance (e.g. $D\left[\rho_{1},\rho_{2}\right]=\tr\abs{\rho_{1}-\rho_{2}}$,
with $\abs A=\sqrt{A^{\dagger}A}$) \cite{Breuer2009a}. For time
independent processes, i.e. $\gamma_{\ell}\left(t\right)=\gamma_{\ell}\geq0$
and $L_{\ell}\left(t\right)=L_{\ell}$, Eq.(\ref{eq:Lindblad}) reduces
to the celebrated Lindblad form \cite{Lindblad1976,Gorini1976a,H.-P.BreuerandF.Petruccione2002}
which can be obtained from the microscopic evolution assuming a small
system-bath coupling and a Markovian (memoryless) environment. The
Markovian assumption has reveled extremely fruitful with the Lindblad
formalism being widely used to model quantum optics and mesoscopic
systems \cite{Carmichael1991,Brandes2005,Vogl2012,Kopylov2013,Eastham2013}
and, more recently, quantum transport \cite{Prosen2008,Prosen2010,Wichterich2007,Medvedyeva2013}.
Master equations of the Lindblad form also allow for efficient stochastic
simulation techniques using Monte-Carlo methods \cite{Plenio1998,H.-P.BreuerandF.Petruccione2002}.
Nonetheless, the evolution of open quantum systems is generically
non-Markovian with some $\gamma_{\ell}$'s assuming negative values.
The Lindblad description fails whenever coherent dynamics between
system and environment are essential. 

If some of the decoherence rates become negative, although $\mathcal{E}_{t,0}$
is completely positive, $\mathcal{E}_{t,t'}$ for $t'>0$ might not
be so. Thus, not all initial density matrices are allowed starting
points for the evolution from $t'$ to $t$, implying that the process
has memory. Non-negative decoherence rates can thus be associated
with memoryless environments \cite{Wolf2008,Breuer2009a,Rivas2010,Hall2014}.
``Non-Markovianity'', i.e. the presence of an environment with a
finite memory time, can be detected and measured using recently proposed
measures and witnesses \cite{Wolf2008,Breuer2009a,Rivas2010,Lorenzo2011,Luo2012,Liu2013,Rivas2014}.
Here, we consider the measure $f_{\text{nM}}\left(t\right)=\frac{1}{2}\sum_{\ell}\left[\abs{\gamma_{\ell}\left(t\right)}-\gamma_{\ell}\left(t\right)\right]$,
strictly quantifying the non-Markovianity \cite{Rivas2010}.

In this letter we explicitly provide a master equation for the class
of quadratic fermionic systems coupled to non-interacting reservoirs.
This extends the knowledge of the exact form of the jump operators
of non-Markovian processes to a wide and important class of models,
used to study spin and electronic transport in normal systems and
superconductors. After providing the explicit form of the jump operators
we show how our results can be applied to treat non-Markovian dynamics
in two examples: a tight-binding model and an open XY-spin chain.
\begin{figure*}
\centering{}\includegraphics[width=1\textwidth]{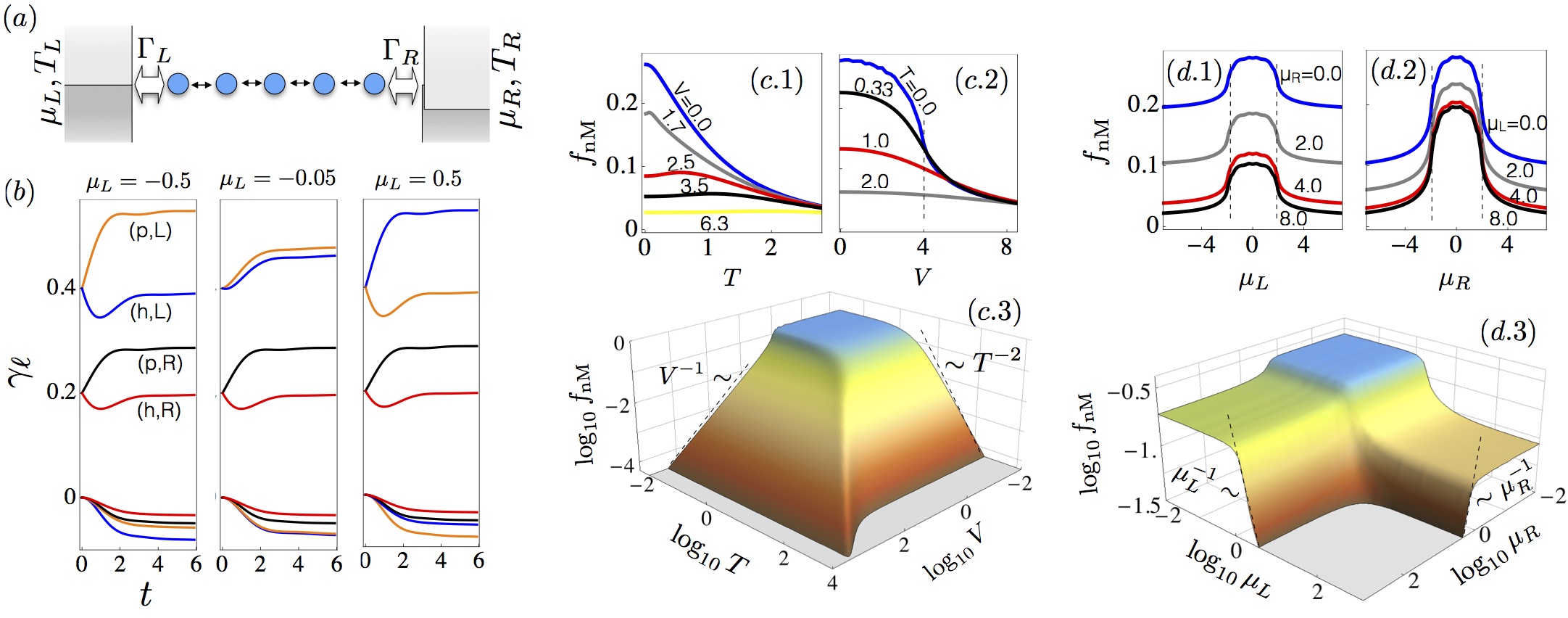}\protect\caption{\label{fig:tight}(a) Sketch of the system coupled to thermal reservoirs.
(b) Decoherence rates $\gamma_{\ell}\left(t\right)$ as a function
of time computed for $M=50,\Gamma_{L}=0.4$, $\Gamma_{R}=0.2,$ $T=0$,
and $\mu_{R}=0.5$ for different values of $\mu_{L}$. The labels
(p/h,L/R) refer to the particle or hole nature of the single-particle
state $\protect\ket{\ell\left(t\right)}$ and to its localization
with respect to the boundary. Negative eigenvalues with the same labels
as their positive counterparts are depicted in the same line-color.
(c) Measure of non-Markovianity $f_{\text{nM}}$ for the steady-state
process computed for $M=50,\Gamma_{L}=0.6$, $\Gamma_{R}=0.2,$ $T_{L}=T_{R}=T$
and $\mu_{L}=-\mu_{R}=V/2$ as a function of $T$ and $V$. (d) The
same as in (c) for $T_{L}=T_{R}=0$ as a function of $\mu_{L}$ and
$\mu_{R}$. }
\end{figure*}

\paragraph{Open quadratic models }

Consider a generic quadratic fermionic system coupled to non-interacting
fermionic reservoirs (leads) labeled by $\nu=1,...,m$. The fermionic
operators of the system and of the reservoirs are denoted $c_{a=1,...,n}$
and $f_{\nu_{a=1,2,...}}$, respectively. The total Hamiltonian is
given by $H=H_{c}+\sum_{\nu}H_{\nu}+H_{c-\nu}$ where $H_{c}=\frac{1}{2}\bs C^{\dagger}\bs H_{c}\bs C$
is the Hamiltonian of the system, with $\bs H_{c}$ the single particle
Hamiltonian and $\boldsymbol{C}=\left\{ c_{1},...,c_{n},c_{1}^{\dagger},....,c_{n}^{\dagger}\right\} ^{T}$
the Nambu vector. $H_{\nu}=\sum_{i}\varepsilon_{\nu}\left(f_{\nu_{i}}^{\dagger}f_{\nu_{i}}-\frac{1}{2}\right)$
is the Hamiltonian of the $\nu$-th reservoir. The interaction Hamiltonian
is given by $H_{c-\nu}=\frac{1}{2}\sum_{\nu}\left(\bs F_{\nu}^{\dagger}\bs T_{\nu}^{\dagger}\bs C+\bs C^{\dagger}\bs T_{\nu}\bs F_{\nu}\right)$
with $\boldsymbol{F}_{\nu}=\left\{ f_{\nu_{1}},...,f_{\nu_{n_{\nu}}},f_{\nu_{1}}^{\dagger},...,f_{\nu_{n_{\nu}}}^{\dagger}\right\} ^{T}$
and $\bs T_{\nu}$ the hopping matrix explicitly given by $\bs T_{\nu}=\sum_{l}\left(t{}_{\nu_{l}}\ket{\nu_{l}}\bra{\Omega_{\nu_{l}}}-\bar{t}_{\nu_{l}}\ket{\hat{\nu_{l}}}\bra{\hat{\Omega}_{\nu_{l}}}\right)$
where $\ket{\nu_{l}}$ is a single-particle state of the system, coupled
to the reservoir $\nu$, $\ket{\Omega_{\nu_{l=1,2,...}}}=\sum_{\varepsilon_{\nu}}\,\Omega_{\nu_{l}}\left(\varepsilon_{\nu}\right)\ket{\varepsilon_{\nu}}$
are single-particle states of the reservoir $\nu$ and $t{}_{\nu_{l}}$
is the hopping amplitude. $\ket{\hat{\nu_{l}}}$ and $\bra{\hat{\Omega}_{\nu_{l}}}$
denote the particle-hole transformed of $\ket{\nu_{l}}$ and $\bra{\Omega_{\nu_{l}}}$. 

After the coupling is turned on at $t=0$, we consider the joint system-reservoir
evolution, taken to be initially in a product state. Each reservoir,
being a macroscopic system, has its initial state specified by $\beta_{\nu}$,
the inverse temperature, and $\mu_{\nu}$, the chemical potential.
The initial density matrix of the system is taken to be of the generic
quadratic form $\rho\left(0\right)=e^{-\frac{1}{2}\bs C^{\dagger}\bs{\Omega}_{0}\bs C}/Z$
with $\bs{\Omega}_{0}$ a single-particle operator. 

The Dyson equation on the Keldysh contour is derived by standard non-equilibrium
Green's functions techniques \cite{Kamenev2011} (the derivation is
sketched in the supplementary material for completeness). At this
point we make a crucial assumption respecting the environment properties
- the so called wide-band limit - which amounts to say that the density
of states $\rho_{\nu}\left(\varepsilon\right)=\sum_{\varepsilon_{\nu}}\delta\left(\varepsilon-\varepsilon_{\nu}\right)$
of the reservoirs and the amplitudes $\Omega_{\nu_{l}}\left(\varepsilon_{\nu}\right)$
are essentially constant with respect to the energy scales of the
system, i.e. $\rho_{\nu}\left(\varepsilon\right)\simeq\rho_{\nu}$,
$\Omega_{\nu_{l}}\left(\varepsilon_{\nu}\right)\simeq\Omega_{\nu_{l}}$.
Denoting $\bs g_{\nu}^{R/A/K}\left(\omega\right)$, the retarded,
advanced and Keldysh components of the bare Greens Function of the
reservoirs, the wide-band limit translates to $\bra{\Omega_{\nu_{l}}}\bs g_{\nu}^{R/A}\left(\omega\right)\ket{\Omega_{\nu_{l'}}}\simeq\mp i\pi\rho_{\nu}\bar{\Omega}_{\nu_{l}}\Omega_{\nu_{l'}}$
and $\bra{\Omega_{\nu_{l}}}\bs g_{\nu}^{K}\left(\omega\right)\ket{\Omega_{\nu_{l'}}}\simeq-2\pi i\rho_{\nu}\bar{\Omega}_{\nu_{l}}\Omega_{\nu_{l'}}\tanh\left[\beta_{\nu}\left(\omega-\mu_{\nu}\right)\right]$.
In this limit, the self-energy components are $\bs{\Sigma}_{c}^{R/A}(t,t')=\mp i\sum_{\nu}\left(\bs{\Gamma}_{\nu}+\hat{\bs{\Gamma}}_{\nu}\right)\delta\left(t-t'\right)$
and $\bs{\Sigma}_{c}^{K}\left(t,t'\right)=-2i\sum_{\nu}\left[\bs{\Gamma}_{\nu}F_{\nu}\left(t-t'\right)-\hat{\bs{\Gamma}}_{\nu}\bar{F}_{\nu}\left(t-t'\right)\right]$,
where $F_{\nu}\left(t\right)=\int\frac{d\varepsilon}{2\pi}\tanh\left[\beta_{\nu}\left(\varepsilon-\mu_{\nu}\right)\right]e^{-i\varepsilon t}$,
$\bs{\Gamma}_{\nu}=\sum_{ll'}\pi\rho_{\nu}\bar{\Omega}_{\nu_{l}}\Omega_{\nu_{l'}}t_{\nu_{l}}\bar{t}_{\nu_{l'}}\ket{\nu_{l}}\bra{\nu_{l'}}$
. A different set of assumptions leading to a similar $\bs{\Sigma}_{c}^{R/A}$
was used in \cite{Dhar2012} to study steady-state transport. The
retarded and advanced Green's functions are given by $\bs G_{c}^{R}\left(t,t'\right)=-i\Theta\left(t-t'\right)e^{-i\left(t-t'\right)\bs K},$
and $\bs G_{c}^{A}\left(t,t'\right)=\bs G_{c}^{R}\left(t',t\right)^{\dagger}$,
where $\bs K=\bs H_{c}-i\bs{\Gamma}$ and $\bs{\Gamma}=\sum_{\nu}\left(\bs{\Gamma}_{\nu}+\hat{\bs{\Gamma}}_{\nu}\right)$.
The Keldysh component is given by $\bs G_{c}^{K}\left(t,t'\right)=e^{-it\bs K}\bs G_{c}^{K}\left(0,0\right)e^{it'\bs K^{\dagger}}+\int_{0}^{t}dt_{1}\int_{0}^{t'}dt_{2}e^{-i\left(t-t_{1}\right)\bs K}\bs{\Sigma}_{c}^{K}\left(t_{1},t_{2}\right)e^{-i\left(t_{2}-t'\right)\bs K^{\dagger}}$
where $\bs G_{c}^{K}\left(0,0\right)=-i\tanh\left(\bs{\Omega}_{0}\right)$
is determined by the initial condition of the system.

\paragraph{Master equation }

Under the evolution given by Eq.(\ref{eq:Lindblad}), for a quadratic
Hamiltonian and linear jump operators $L_{\ell}\left(t\right)=\sum_{i}\braket{\ell\left(t\right)}iC_{i}$
(with $\braket{\ell\left(t\right)}{\ell'\left(t\right)}=\delta_{\ell,\ell'}$),
an initial Gaussian density matrix remains of the Gaussian form: $\rho\left(t\right)=e^{-\frac{1}{2}\bs C^{\dagger}\bs{\Omega}\left(t\right)\bs C}/Z\left(t\right)$
and the single-particle correlation matrix, given by $\bs{\chi}\left(t\right)=\av{\bs C\left(t\right).\bs C^{\dagger}\left(t\right)}=\left[1+e^{-\bs{\Omega}\left(t\right)}\right]^{-1}$,
fully encodes all the equal-time properties of the system. Under the
Lindblad dynamics $\bs{\chi}\left(t\right)$ evolves as (see \cite{Prosen2008a}
and supplementary material for a derivation):
\begin{eqnarray}
\pd_{t}\bs{\chi}\left(t\right) & = & -i\bs Q\left(t\right)\bs{\chi}\left(t\right)+i\bs{\chi}\left(t\right)\bs Q^{\dagger}\left(t\right)+\bs N\left(t\right)\label{eq:evol_chi}
\end{eqnarray}
with $\bs N\left(t\right)=\sum_{\ell}\gamma_{\ell}\left(t\right)\ket{\ell\left(t\right)}\bra{\ell\left(t\right)}$
and $\bs Q\left(t\right)=\bs H_{c}\left(t\right)-i\frac{1}{2}\left[\bs N\left(t\right)+\hat{\bs N}\left(t\right)\right]$.
Using $\bs{\chi}\left(t\right)=\frac{1}{2}\left[i\bs G_{c}^{K}\left(t,t\right)+1\right]$
and deriving in order to $t$, we can identify the different elements
of Eq.(\ref{eq:evol_chi}):
\begin{eqnarray}
\bs Q\left(t\right)=\bs K; & \ \, & \bs N\left(t\right)=\sum_{\nu}\bs N_{\nu}\left(t\right)\label{eq:Q_K}
\end{eqnarray}
with
\begin{multline}
\bs N_{\nu}\left(t\right)=\left(\bs{\Gamma}_{\nu}+\hat{\bs{\Gamma}}_{\nu}\right)\\
+i\left\{ R\left[\left(\bs K+\mu_{\nu}\right),\beta_{\nu},t\right]\bs{\Gamma}_{\nu}-\bs{\Gamma}_{\nu}R\left[\left(\bs K+\mu_{\nu}\right),\beta_{\nu},t\right]^{\dagger}\right.\\
\left.+R\left[\left(\bs K-\mu_{\nu}\right),\beta_{\nu},t\right]\hat{\bs{\Gamma}}_{\nu}-\hat{\bs{\Gamma}}_{\nu}R\left[\left(\bs K-\mu_{\nu}\right),\beta_{\nu},t\right]^{\dagger}\right\} \label{eq:N_nu}
\end{multline}
where $R\left[\omega,\beta,t\right]=s\left[\beta\omega,t/\beta\right]+r\left[\omega t\right]$,
with $s\left[z,\tau\right]=-\int_{0}^{\tau}d\tau'\, e^{-iz\tau'}\int_{0}^{\infty}dx\left(\tanh\left[x\right]-1\right)\sin\left(x\tau'\right)/\pi$
and $r\left[x\right]=\left[\log\left(ix\right)+\Gamma\left(0,ix\right)+\gamma+\log\left(\frac{4}{\pi}\right)\right]/\pi$,
are obtained by a suitable regularization of the wide-band limit (see
supplementary material). 

The decoherence rates $\gamma_{\ell}\left(t\right)$ and the vectors
$\ket{\ell\left(t\right)}$, characterizing the jump operators, can
be obtained by diagonalizing $\bs N\left(t\right)$. Eqs. (\ref{eq:Q_K})
show explicitly how to obtain the master equation describing a non-Markovian
process and are the central result of this letter. The more general
case where the system Hamiltonian and the system-environment couplings
depend on time is straitforwardly obtained and is given in the supplementary
material. 

Particularly simple cases yielding to the Markovian dynamics arise
for fully empty or fully filled reservoirs \cite{Gurvitz1996,Gurvitz1998},
i.e. $\mu_{\nu}\to\pm\infty$, for which $\bs N_{\nu}\left(t\right)=2\hat{\bs{\Gamma}}_{\nu}$
and $\bs N_{\nu}\left(t\right)=2\bs{\Gamma}_{\nu}$ respectively,
and for the infinite temperature, $\beta_{\nu}\to0$, for which $\bs N_{\nu}\left(t\right)=\bs{\Gamma}_{\nu}+\hat{\bs{\Gamma}}_{\nu}$. 

In the asymptotic long time limit $\bs N\left(t\right)$ converges
to a time-independent matrix $\bs N_{\infty}$. If a unique steady-state
exists, the single particle density matrix is given by $\bs{\chi}_{\infty}=-i\sum_{\beta\gamma}\ket{\beta}\frac{\bra{\beta'}\bs N_{\infty}\ket{\gamma'}}{\lambda_{\beta}-\bar{\lambda}_{\gamma}}\bra{\gamma}$
where $\ket{\beta}$ and $\bra{\beta'}$ are right and left eigenvectors
of $\bs K$ with eigenvalue $\lambda_{\beta}$ and $\braket{\gamma}{\beta'}=\delta_{\gamma\beta}$.
\begin{figure}
\centering{}\includegraphics[width=1\columnwidth]{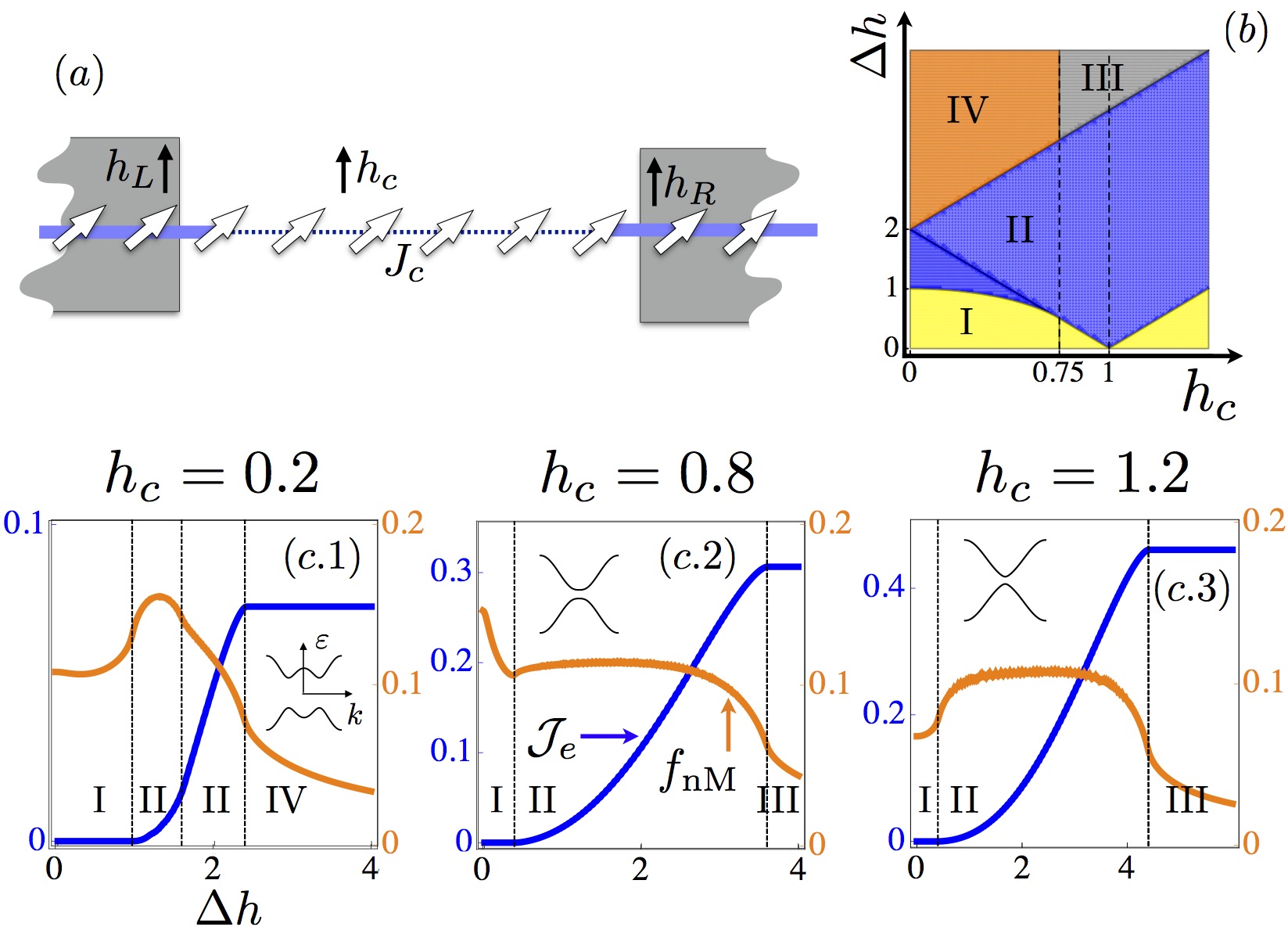}\protect\caption{\label{fig:XY}(a) Sketch of the XY model coupled to spin reservoirs
with $T_{L}=T_{R}=0$ and $h_{L}=-h_{R}=\Delta h$. (b) Phase diagram
of the non-equilibrium steady-state in the $h_{c}-\mathbf{\Delta}h$
plane computed for $\gamma_{c}=0.5$, $J_{c}=1$. Regions I to IV
are described in the text. (c) Measure of non-Markovianity $f_{\text{nM}}$
and energy current $\mathcal{J}_{e}$ as a function of the spin unbalance
$\Delta h$ computed for different values of $h_{c}$ and $\gamma_{c}=0.5$.
The band-structure of the spin-less Jordan-Wigner fermions are depicted
in the insets. }
\end{figure}

\paragraph{Tight-binding chain}

In order to demonstrate our approach let us consider a tight-binding
one-dimensional chain in Fig.\ref{fig:tight}-(a), with $\bs H_{c}=\text{diag}\left(\bs h,-\bs h^{T}\right)$
and $\bs h=-\sum_{j=0}^{M-2}\ket j\bra{j+1}+\ket{j+1}\bra j$, coupled
to two leads at positions $0$ and $M-1$ by the hybridization operators
$\bs{\Gamma}_{L}=\Gamma_{L}\ket 0\bra 0$ and $\bs{\Gamma}_{R}=\Gamma_{R}\ket{M-1}\bra{M-1}$. 

Fig.\ref{fig:tight}-(b) shows the evolution of the decoherence rates
$\gamma_{\ell}\left(t\right)$, after the coupling to the reservoirs
has been turned on, for different values of $\mu_{L}$. There are
8 non-zero eigenvalues of $\bs N$, arising in positive-negative pairs
(see code color). In the Markovian limit the negative eigenvalues
tend to zero. The labels p/h refer to the particle or hole nature
of the corresponding eigenvector of $\bs N$, and L/R to their localization
near the left or right lead. For $M\to\infty$ we observe that $\bs N_{R/L}\ket{\ell_{L/R}\left(t\right)}\to0$,
where $\bs N_{\nu}\left(t\right)\ket{\ell_{\nu}\left(t\right)}=\gamma_{\ell;\nu}\left(t\right)\ket{\ell_{\nu}\left(t\right)}$,
i.e. for a large size chain the contribution of both reservoirs factorizes
and the non-zero eigenvalues of $\bs N$ can be obtained by direct
sum of the spectrum of $\bs N_{L}$ and $\bs N_{R}$. This factorization
explains that in Fig.\ref{fig:tight}-(b) the R-labeled eigenvalues
are unaffected by changes in $\mu_{L}$. More generally, such a factorization,
arising when the special separation between the reservoirs is large,
is to be expected for short-range Hamiltonians $H_{c}$ and allows
to treat the decoherence rates of each reservoir independently. In
the present example the structure of $\bs N_{\nu}$ is particularly
simple: $\bs N_{\nu}\left(t\right)=\ket{y_{\nu}^{p}\left(t\right)}\bra{\nu}+\ket{\nu}\bra{y_{\nu}^{p}\left(t\right)}+\ket{y_{\nu}^{h}\left(t\right)}\bra{\hat{\nu}}+\ket{\hat{\nu}}\bra{y_{\nu}^{h}\left(t\right)}$
with $\ket{y_{\nu}^{p}\left(t\right)}=\left\{ \frac{1}{2}+iR\left[\left(\bs K+\mu_{\nu}\right),\beta_{\nu},t\right]\right\} \bs{\Gamma}_{\nu}\ket{\nu}$
and $\ket{y_{\nu}^{h}\left(t\right)}=\left\{ \frac{1}{2}+iR\left[\left(\bs K-\mu_{\nu}\right),\beta_{\nu},t\right]\right\} \hat{\bs{\Gamma}}_{\nu}\ket{\hat{\nu}}$;
yielding to $\gamma_{\left(p,\nu\right)}^{\pm}=\frac{1}{2}\left\{ \braket{\nu}{y_{\nu}^{p}}+\frac{1}{2}\braket{y_{\nu}^{p}}{\nu}\right.\pm\left(\left(\braket{\nu}{y_{\nu}^{p}}+\braket{y_{\nu}^{p}}{\nu}\right)^{2}\right.+\left.\left.4\left(\braket{y_{\nu}^{p}}{y_{\nu}^{p}}-\braket{\nu}{y_{\nu}^{p}}\braket{y_{\nu}^{p}}{\nu}\right)\right)^{1/2}\right\} $
and to a similar expression for their hole counterparts, corresponding
to the two positive and negative eigenvalue pairs in Fig.\ref{fig:tight}-(b).
Note that $\gamma_{\left(p,\nu\right)}^{-}$ is zero only (Markovian
case) if $\ket{y_{\nu}^{p}}\propto\ket{\nu}$. The fact that in Fig.\ref{fig:tight}-(b)
the particle or hole nature of the L-labeled eigenvalues is interchanged
upon switching $\mu_{L}\to-\mu_{L}$ can be seen in the expressions
of $\ket{y_{\nu}^{p/h}}$ together with the fact that $\bs K$ has
no anomalous terms. 

Figs.\ref{fig:tight}-(c) and (d) depict the non-Markovianity nature
of the steady-state as measured by the $f_{\text{nM}}=f_{\text{nM}}\left(t\to\infty\right)$.
In Figs.\ref{fig:tight}-(c.1,2,3) we set $\mu_{L}=-\mu_{R}=V/2$;
\textbf{$T_{L}=T_{R}=T$} and show $f_{\text{nM}}$ as a function
of the bias voltage $V$ and temperature $T$. Figs.\ref{fig:tight}-(c.1)
and (c.2) show how $f_{\text{nM}}$ varies as a function of $T$ and
$V$ respectively. Fig.\ref{fig:tight}-(c.3) shows a logarithmic
plot of $f_{\text{nM}}$ for large values of $V$ and $T$. The Markovian
limit, obtained for large values $T$ or $V$, is attained differently
along the two axes: $f_{\text{nM}}\propto V^{-1}$ for large $V$
and $f_{\text{nM}}\propto T^{-2}$ for large $T$. 

Figs.\ref{fig:tight}-(d) shows the variation of $f_{\text{nM}}$
with $\mu_{L}$ and $\mu_{R}$ separately at \textbf{$T_{L}=T_{R}=0$}.
Fig.\ref{fig:tight}-(d.3) shows clearly that the Markovian limit
is attained only when both chemical potentials are large. This can
be understood by the approximate factorization of the eigenvalues
of $\bs N$ as a Markovian evolution can only arise when both reservoirs
behave as memoryless environments. For $\abs{\mu_{L/R}}\gg\abs{\mu_{R/L}}$
one has $f_{\text{nM}}\propto\abs{\mu_{L/R}}^{-1}$.

\paragraph{XY spin-chain}

In the Markovian limit a number of works have addressed spin and heat
transport in spin-chains \cite{Wichterich2007,Prosen2008,Prosen2010,Vogl2012,Cai2013,Ajisaka2014}.
Here, we consider a XY spin-chain with non-Markovian reservoirs, depicted
in Fig.\ref{fig:XY}-(a). The Hamiltonian is given by $H=-\sum_{m}\frac{J}{2}\left[\left(1+\gamma\right)\sigma_{m}^{x}\sigma_{m+1}^{x}\right.+\left.\left(1-\gamma\right)\sigma_{m}^{y}\sigma_{m+1}^{y}\right]-h\sum_{m}\sigma_{m}^{z},$
where $J=J_{c}$, $\gamma=\gamma_{c}$ and $h=h_{c}$ within the central
region. Setting $J=J_{L/R}$ with $J_{L/R}/J_{c}\gg1$ and $\gamma=0$,
the side chains act as wide-band gapless reservoirs with $h=h_{L/R}$.
In the following we set $h_{L}=-h_{R}=\Delta h$ and work in units
where $J_{c}=1$. The coupling Hamiltonian is given by $H_{int}=-\frac{J'_{L}}{2}\left[\sigma_{L,0}^{x}\sigma_{0}^{x}+\sigma_{L,0}^{y}\sigma_{0}^{y}\right]-\frac{J'_{R}}{2}\left[\sigma_{R,0}^{x}\sigma_{M-1}^{x}+\sigma_{R,0}^{y}\sigma_{M-1}^{y}\right]$.
Employing a Jordan-Wigner mapping this model can be transformed into
a set of non-interacting spineless fermions. For the central region
one has $\bs H_{c}=\left(\begin{array}{cc}
\bs h & \bs{\Delta}\\
\bs{\Delta}^{\dagger} & \bs{-h}^{T}
\end{array}\right)$ with $\bs h=-J_{c}\sum_{j=0}^{M-2}(\ket m\bra{m+1}+\ket{m+1}\bra m)-2h_{c}\sum_{m=0}^{M-1}\ket m\bra m$
and $\bs{\Delta}=J_{c}\gamma_{c}\sum_{j=0}^{M-2}\left[\ket m\bra{\hat{m+1}}-\ket{m+1}\bra{\hat{m}}\right]$.
Following our wide-band treatment for the reservoirs (i.e. $J_{L/R}/J_{c}\to\infty$)
we obtain $\bs{\Gamma}_{L}=\Gamma_{L}\ket 0\bra 0$ and $\bs{\Gamma}_{R}=\Gamma_{R}\ket{M-1}\bra{M-1}$,
where $\Gamma_{L/R}\propto J'^{2}/J_{L/R}$ are constants that characterize
the contacts, and $\mu_{L/R}=2h_{L/R}$. 

In the Markovian limit ($\Delta h\to\infty$) this model was shown
to exhibit a steady-state phase transition, where the decay of the
correlators $C_{l,m}=\av{\sigma_{l}^{z}\sigma_{m}^{z}}-\av{\sigma_{l}^{z}}\av{\sigma_{m}^{z}}$,
as a function of $r=\abs{l-m}$, passes from power law (for $h_{c}/J_{c}<1-\gamma_{c}^{2}$)
to exponential (for $h_{c}/J_{c}>1-\gamma_{c}^{2}$) \cite{Prosen2008}.
We address the non-Markovian regime (finite $\Delta h$) and monitor
the steady-state energy-current $\mathcal{J}_{e}$ and $f_{\text{nM}}$
in addition to $C_{l,m}$ (the explicit forms of $\mathcal{J}_{e}$
and $C_{l,m}$ are given in the supplementary material). Fig.\ref{fig:XY}-(b)
shows the phase diagram in the $h_{c}-\Delta h$ plane and signals
the four different steady-state phases. The energy current and $f_{\text{nM}}$
as a function of $\Delta h$ are given in Figs.\ref{fig:XY}-(c.1-3)
for different values of $h_{c}.$ A numerical demonstration of the
exponential/algebraic decay of $C_{l,m}$ within each region is provided
in the supplementary material. In region I both effective chemical
potentials ($\mu_{L/R}$) are below the excitation-gap. This region
shows a vanishing energy current and an exponential decay of $C_{l,m}$.
In region II there is energy transport with a finite $d\mathcal{J}_{e}/d\Delta h$
and an algebraic decay of $C_{l,m}$. In this region $\mu_{R/L}$
lay within the excitation energy band. Region III and IV show a saturation
of the energy current and $f_{\text{nM}}$ behaves as $1/\Delta h$
as the Markovian limit is taken. However, in III, $C_{l,m}$ is algebraically
decaying whereas is IV the decay is exponential. 

These results show that the two Markovian phases reported in \cite{Prosen2008}
can be continuously connected to phases III and IV. Moreover deep
into the non-Markovian regime phases I and II arise having no non-Markovian
analog.

\paragraph{Discussion}

We provide an explicit construction of the master equations for quadratic
fermionic models coupled to wide-band reservoirs by identifying the
jump operators and the decoherence rates derived with the non-equilibrium
Green's functions formalism. This approach permits to study non-Markovian
regimes characterized by negative decoherence rates and to clarify
the regimes where the Markovian approximation yields a good approximation
for the dynamics. We illustrate our findings with two examples of
non-Markonian evolution. The XY model shows a particularly rich set
of phases with distinct physical properties. 

Our results provide an explicit approach to study real-time dynamics
of a wide class of open systems. As quadratic models are often used
as starting points of perturbative and variational approaches, our
results might also be of interest to study master-equations of interacting
models.
\begin{acknowledgments}
During part of this work PR was supported by the Marie Curie International
Reintegration Grant PIRG07-GA-2010-268172. 
\end{acknowledgments}

\bibliographystyle{apsrev4-1}
\bibliography{VRVieira}

\appendix
\clearpage{}

\begin{widetext}

\part*{Supplementary Material}

\section{Preliminary considerations }

\subsection{Notation }

For a generic fermionic system with $n$ modes, obeying the anti-commutation
relations $\left\{ c_{a},c_{b}^{\dagger}\right\} =\delta_{a,b};\,\left\{ c_{a},c_{b}\right\} =\left\{ c_{a}^{\dagger},c_{b}^{\dagger}\right\} =0$,
$\left(a,b=1,...,n\right)$, we define $\boldsymbol{C}=\left\{ c_{1},...,c_{n},c_{1}^{\dagger},....,c_{n}^{\dagger}\right\} ^{T}$
as the column vector of annihilation and creation operators. For definiteness
we take the indices $a$ and $b$ as labeling the position of a fermion
on a finite lattice, such that $\ket a$ ($\ket{\hat{a}}$ ) corresponds
to a particle (hole) at position $a$. The indices $i,j$ are used
to label all single-particle or hole states $\ket i\in\left\{ \ket a\right\} \cup\left\{ \ket{\hat{a}}\right\} $.
With these definitions one has $\bra a\boldsymbol{C}=c_{a}$, $\bra{\hat{a}}\boldsymbol{C}=c_{a}^{\dagger}$
or equivalently $\bra i\bs C=\bs C_{i}$. In the following, the bold
symbols are used for $2n\times2n$ matrices and $2n$ vectors. In
this way a generic single-body operator $A=\sum_{a,b}(c_{a}^{\dagger}A_{a,b}^{pp}c_{a}+c_{a}A_{a,b}^{hh}c_{a}^{\dagger}+c_{a}^{\dagger}A_{a,b}^{ph}c_{b}^{\dagger}+c_{a}A_{a,b}^{hp}c_{b})$
can be written as $A=\bs C^{\dagger}\bs A\bs C$ with $\mathbf{A}=\sum_{a,b}(\ket aA_{a,b}^{pp}\bra b+\ket aA_{a,b}^{ph}\bra{\hat{b}}+\ket{\hat{a}}A_{a,b}^{hp}\bra b+\ket{\hat{a}}A_{a,b}^{hh}\bra{\hat{b}})=\sum_{ij}\ket i\mathbf{A}_{i,j}\bra j$
and $\tr\bs A=0$. We define the particle-hole transform of a single-particle
state $\ket{\phi}=\sum_{a}(\phi_{a}^{p}\ket a+\phi_{a}^{h}\ket{\hat{a}})=\sum_{i}\phi_{i}\ket i$
as $\ket{\hat{\phi}}=\bs J\ket{\phi^{*}}$, where the conjugate $\ket{\phi^{*}}=\sum_{i}\bar{\phi}_{i}\ket i$
is taken with respect to the basis $\ket i$ and $\bs J=\sum_{a}\left(\ket a\bra{\hat{a}}+\ket{\hat{a}}\bra a\right)$
transforms single-particle (hole) states into their hole (particle)
analog. A similar definition holds for the operators $\hat{\bs A}=\bs J\bs A^{T}\bs J$,
with $\bs A^{T}=\sum_{ij}\ket i\mathbf{A}_{j,i}\bra j$.

\subsection{Green's functions and single-body density matrix}

We define the greater and lesser Green's functions, containing both
normal (i.e. $c_{a}^{\dagger}c_{b}$ and $c_{a}c_{b}^{\dagger}$ )
and anomalous (i.e. $c_{a}^{\dagger}c_{b}^{\dagger}$ and $c_{a}c_{b}$)
terms, as 
\begin{eqnarray}
\mathbf{G}_{i,j}^{>}\left(t,t'\right) & \equiv & -i\av{\bs C_{i}\left(t\right)\bs C_{j}^{\dagger}\left(t'\right)}\\
\mathbf{G}_{i,j}^{<}\left(t,t'\right) & \equiv & i\av{\bs C_{j}^{\dagger}\left(t'\right)\bs C_{i}\left(t\right)}\label{eq:GF_def}
\end{eqnarray}
The retarded, advanced and Keldysh Green's functions are defined in
the standard way 
\begin{eqnarray}
\mathbf{G}^{R}(t,t') & \equiv & \Theta(t-t')\left[\mathbf{G}^{>}(t,t')-\mathbf{G}^{<}(t,t')\right]\\
\mathbf{G}^{A}(t,t') & \equiv & -\Theta(t'-t)\left[\mathbf{G}^{>}(t,t')-\mathbf{G}^{<}(t,t')\right]\\
\mathbf{G}^{K}(t,t') & \equiv & \mathbf{G}^{>}(t,t')+\mathbf{G}^{<}(t,t')\label{eq:GF_ARK}
\end{eqnarray}
The single-body correlation matrix $\bs{\chi}=\av{\bs C\bs C^{\dagger}}$
can be obtained as the equal time limit of the greater Green's function
\begin{eqnarray}
\bs{\chi}_{ij}\left(t\right) & = & \av{\bs C_{i}\left(t\right)\bs C_{j}^{\dagger}\left(t\right)}=i\bs G_{i,j}^{>}(t,t)
\end{eqnarray}
Noting that the greater Green's function can be obtained as $\bs G^{>}=\frac{1}{2}\left[\bs G_{c}^{K}+\bs G_{c}^{R}-\bs G_{c}^{A}\right]$
and $\bs G^{R}\left(t,t\right)-\bs G^{A}\left(t,t\right)=-i$ this
quantity is simply related to the Keldysh Green's function
\begin{eqnarray}
\bs{\chi}\left(t\right) & = & \frac{1}{2}\left[i\bs G^{K}\left(t,t\right)+1\right].\label{eq:chi_GK}
\end{eqnarray}
$\bs{\chi}\left(t\right)$ has the information about all equal time
single-body correlations, for example: $\av{c_{a}^{\dagger}\left(t\right)c_{b}\left(t\right)}=\bra{\hat{a}}\bs{\chi}\left(t\right)\ket{\hat{b}}$.
From the commutation relations among fermions and the definition of
particle hole symmetry, $\bs{\chi}$ respects: 
\begin{eqnarray}
\bs{\chi}^{\dagger} & = & \bs{\chi}\\
\tr\left(\bs{\chi}\right) & = & n\\
\hat{\bs{\chi}} & = & 1-\bs{\chi}
\end{eqnarray}

\subsection{Closed quadratic models\label{sub:Generic-considerations-for}}

A generic quadratic Hamiltonian can be written as
\begin{eqnarray}
H & = & \frac{1}{2}\bs C^{\dagger}\bs H\bs C
\end{eqnarray}
where $\bs{H=}\bs H^{\dagger}$ is the single-body Hamiltonian given
by
\begin{eqnarray}
\bs H & = & \sum_{a,b}\left(\ket ah_{ab}\bra b-\ket{\hat{a}}h_{ba}\bra{\hat{b}}+\ket a\Delta_{ab}\bra{\hat{b}}+\ket{\hat{a}}\bar{\Delta}_{ba}\bra b\right)\label{eq:H_generic}
\end{eqnarray}
where $h$ and $\Delta$ are $n\times n$ matrices with the properties
$h^{\dagger}=h$ and $\Delta^{T}=-\Delta$. Note that $\bs H$ fulfills
the particle-hole conjugation condition $\hat{\bs H}=-\bs H$ implying
that if $\bs H\ket{\varepsilon}=\varepsilon\ket{\varepsilon}$ then
$\bs H\ket{\hat{\varepsilon}}=-\varepsilon\ket{\hat{\varepsilon}}$. 

For a non-interacting fermionic system in thermal equilibrium at $t=0$
with the Hamiltonian $H_{0}=\frac{1}{2}\bs C^{\dagger}\bs H_{0}\bs C$,
temperature $k_{B}T=\beta^{-1}$ and chemical potential $\mu$, the
density matrix is given by $\rho=e^{-\beta\left(H-\mu N\right)}/Z$
with $Z=\tr\left[e^{-\beta\left(H-\mu N\right)}\right]$. Evolving
the equilibrium condition under the Hamiltonian $H\left(t\right)=\frac{1}{2}\bs C^{\dagger}\bs H\left(t\right)\bs C$,
the Green's functions in Eq.(\ref{eq:GF_def}) are explicitly given
by 
\begin{eqnarray}
\mathbf{G}^{>}\left(t,t'\right) & = & -i\bs U\left(t,0\right)\left[1-n_{f}\left(\bs H_{0}-\mu\bs N\right)\right]\bs U\left(0,t'\right)\\
\mathbf{G}^{<}\left(t,t'\right) & = & i\bs U\left(t,0\right)n_{f}\left(\bs H_{0}-\mu\bs N\right)\bs U\left(0,t'\right)
\end{eqnarray}
where $\bs U\left(t,t'\right)=Te^{-i\int_{t'}^{t}d\tau\bs H\left(\tau\right)}$
is the single-body evolution operator with $T$ the time ordering
operator, $n_{f}\left(z\right)=\frac{1}{e^{\beta z}+1}$ the Fermi-function
and $\bs N=\sum_{a}\ket a\bra a-\ket{\hat{a}}\bra{\hat{a}}$ corresponds
to the second quantized operator $N=\frac{1}{2}\bs C^{\dagger}\bs N\bs C+\frac{1}{2}n$
that counts the total number of particles in the system. 

For the particular case of time independent Hamiltonian $H\left(t\right)=H$,
the Green's functions in Eq.(\ref{eq:GF_ARK}) become 
\begin{eqnarray}
\mathbf{G}^{R}(t,t') & = & -i\Theta(t-t')e^{-i\bs H\left(t-t'\right)}\\
\mathbf{G}^{A}(t,t') & = & i\Theta(t-t')e^{-i\bs H\left(t-t'\right)}\\
\mathbf{G}^{K}(t,t') & = & -ie^{-i\bs Ht}\left[1-2n_{f}\left(\bs H_{0}-\mu\bs N\right)\right]e^{i\bs Ht'}
\end{eqnarray}
Moreover, if $H_{0}=H$ and $H$ conserves the number of particles
$\left[H,N\right]=0$, all these quantities depend on the difference
of times only: $\mathbf{G}^{R,A,K}(t,t')=\mathbf{G}^{R,A,K}(t-t')$,
and thus 
\begin{eqnarray}
\mathbf{G}^{R/A}(\omega) & = & \left(\omega-\bs H\pm i\eta\right)^{-1}\\
\mathbf{G}^{K}(\omega) & = & -2\pi i\left[1-2n_{f}\left(\bs H-\mu\bs N\right)\right]\delta\left(\omega-\bs H\right)
\end{eqnarray}
with $\mathbf{G}^{R,A,K}\left(\omega\right)=\int dt\, e^{i\omega t}\mathbf{G}^{R,A,K}\left(t\right)$.

\subsection{Derivation of Dyson's equation on the Keldysh contour \label{sec:Derivation-of-Dyson's}}

Consider the generating function on the Keldysh contour, 
\begin{eqnarray}
Z\left[\bs{\eta},\bs{\eta}'\right] & = & \int DcDf\, e^{i\int_{\gamma}dz\,\frac{1}{2}\bs{\psi}^{\dagger}\left(z\right)\left[i\bs{\pd}_{z}-\bs H\right]\bs{\psi}\left(z\right)}e^{\int_{\gamma}dz\left[\bs{\eta}^{\dagger}\left(z\right).\bs{\psi}\left(z\right)+\bs{\psi}^{\dagger}\left(z\right).\bs{\eta}'\left(z\right)\right]}
\end{eqnarray}
where $\bs{\psi}=\left(\bs C,\bs F_{\nu_{1}},\bs F_{\nu_{2}},...\right)$,
$\bs{\eta}$ and $\bs{\eta}'$ are Grassmanian sources and where the
single-particle Hamiltonian $\bs H$ is given by 
\begin{eqnarray}
\bs H & = & \left(\begin{array}{cccc}
\bs H_{C} & \bs T_{\nu_{1}} & \bs T_{\nu_{2}} & \ldots\\
\bs T_{\nu_{1}}^{\dagger} & \bs H_{\nu_{1}} & 0 & \cdots\\
\bs T_{\nu_{2}}^{\dagger} & 0 & \bs H_{\nu_{2}} & \ddots\\
\vdots & \vdots & \ddots & \ddots
\end{array}\right).
\end{eqnarray}
Integrating out the fermions yields to 
\begin{eqnarray}
\frac{Z\left[\bs{\eta},\bs{\eta}'\right]}{Z\left[\bs 0,\bs 0\right]} & = & e^{\frac{i}{2}\int_{\gamma}dzdz'\left(\bs{\eta}^{\dagger}-\bs{\eta}'^{\dagger}\right)\left(z\right)\bs G\left(z,z'\right)\left(\bs{\eta}'-\bs{\eta}\right)\left(z'\right)}\label{eq:Gen_F-1}
\end{eqnarray}
with 
\begin{eqnarray}
\bs G & = & \left(\begin{array}{cc}
\bs G_{cc} & \bs G_{cf_{\nu}}\\
\bs G_{f_{\nu}c} & \bs G_{f_{\nu}f_{\nu'}}
\end{array}\right)
\end{eqnarray}
and 
\begin{eqnarray}
\bs G_{cc} & = & \left[\bs g_{c}^{-1}-\bs{\Sigma}_{c}\right]^{-1}\\
\bs G_{f_{\nu}c} & = & \bs g_{f_{\nu}}\bs T_{\nu}^{\dagger}\bs G_{cc}\\
\bs G_{cf_{\nu}} & = & \bs G_{cc}\bs T_{\nu}\bs g_{f_{\nu}}\\
\bs G_{f_{\nu}f_{\nu'}} & = & \delta_{\nu\nu'}\bs g_{f_{\nu}}+\bs g_{f_{\nu}}\bs T_{\nu}^{\dagger}\bs G_{cc}\bs T_{\nu'}\bs g_{f_{\nu'}}
\end{eqnarray}
where 
\begin{eqnarray}
\bs{\Sigma}_{c} & = & \sum_{\nu}\bs T_{\nu}\bs g_{\nu}\bs T_{\nu}^{\dagger}
\end{eqnarray}
Deriving both sides of Eq.(\ref{eq:Gen_F-1}) in order to the sources
we can verify that $\left[\bs G_{ab}\right]_{i,j}\left(t,t'\right)=\bs G_{a_{i}b_{j}}\left(t,t'\right)$
are the path ordered Green's function $\bs G_{a_{i}b_{j}}\left(t,t'\right)=-i\av{T_{\gamma}a_{i}\left(t\right)b_{j}^{\dagger}\left(t'\right)}$
and where $T_{\gamma}$ is the path ordering operator on the Keldysh
contour. $\bs g_{f_{\nu}}$ and $\bs g_{c}$ are the bare Green's
functions of lead $\nu$ and of the system respectively.

\section{System self-energy}

\subsection{Self-energy}

Using the results derived for closed quadratic models, the retarded,
advanced and Keldysh Green's functions of the reservoirs, in frequency
domain, are given by 
\begin{eqnarray}
\bs g_{\nu}^{R/A}\left(\omega\right) & = & \sum_{\varepsilon_{\nu}}\left(\ket{\varepsilon_{\nu}}\frac{1}{\omega-\varepsilon_{\nu}\pm i\eta}\bra{\varepsilon_{\nu}}+\ket{\hat{\varepsilon}_{\nu}}\frac{1}{\omega+\varepsilon_{\nu}\pm i\eta}\bra{\hat{\varepsilon}_{\nu}}\right)\label{eq:g_K}\\
\bs g_{\nu}^{K}\left(\omega\right) & = & -2\pi i\sum_{\varepsilon_{\nu}}F_{\nu}\left(\omega\right)\left(\ket{\varepsilon_{\nu}}\delta\left(\omega-\varepsilon_{\nu}\right)\bra{\varepsilon_{\nu}}-\ket{\hat{\varepsilon}_{\nu}}\delta\left(\omega+\varepsilon_{\nu}\right)\bra{\hat{\varepsilon}_{\nu}}\right)
\end{eqnarray}
with $F_{\nu}\left(\omega\right)=\left[1-2\frac{1}{e^{\beta_{\nu}\left(\omega-\mu_{\nu}\right)}+1}\right]$.
Using the Langreth's rules we can then obtain the retarded, advanced
and Keldysh components of the system's self-energy, due to the presence
of the reservoirs: 
\begin{eqnarray*}
\bs{\Sigma}_{c}^{R/A/K}(t,t') & = & \sum_{\nu,l}\left(t{}_{\nu_{l}}\bar{t}{}_{\nu_{l}'}\ket{\nu_{l}}\bra{\Omega_{\nu}}\bs g_{\nu}^{R/A/K}\left(t,t'\right)\ket{\Omega_{\nu}}\bra{\nu_{l'}}\right.\\
 &  & \left.+\bar{t}_{\nu_{l}}t_{\nu_{l'}}\ket{\hat{\nu}_{l}}\bra{\hat{\Omega}_{\nu_{l}}}\bs g_{\nu}^{R/A/K}\left(t,t'\right)\ket{\hat{\Omega}_{\nu_{l'}}}\bra{\nu_{l'}}\right).
\end{eqnarray*}

\subsection{Green's functions}

\subsubsection{Properties of operators}

To treat the generic time dependent case, we are going to assume in
this section that the system Hamiltonian $\bs H_{c}\left(t\right)$,
the hopping amplitudes $t_{\nu_{l}}\left(t\right)$ and the single-particle
states $\ket{\nu_{l}\left(t\right)}$ depend on time. In this way
the matrices $\bs{\Gamma}_{\nu}$ in the main text generalize to 

\begin{eqnarray}
\bs{\Gamma}_{\nu}\left(t,t'\right) & = & \pi\rho_{\nu}\left(0\right)\sum_{ll'}\bar{\Omega}_{\nu_{l}}\left(0\right)\Omega_{\nu_{l'}}\left(0\right)t_{\nu_{l}}\left(t\right)\bar{t}_{\nu_{l'}}\left(t'\right)\ket{\nu_{l}\left(t\right)}\bra{\nu_{l'}\left(t'\right)}.
\end{eqnarray}
It is easy to see that:
\begin{eqnarray*}
\left[\bs{\Gamma}_{\nu}\left(t,t'\right)\right]^{\dagger} & = & \bs{\Gamma}_{\nu}\left(t',t\right)
\end{eqnarray*}
Defining $\bs{\Gamma}\left(t\right)=\sum_{\nu}\left[\bs{\Gamma}_{\nu}\left(t,t\right)+\hat{\bs{\Gamma}}_{\nu}\left(t,t\right)\right]$
and $\bs K\left(t\right)=\bs H\left(t\right)-i\bs{\Gamma}\left(t\right)$,
the particle-hole symmetric transformation yields 
\begin{eqnarray*}
\hat{\bs H}\left(t\right) & = & -\bs H\left(t\right)\\
\hat{\bs{\Gamma}}\left(t\right) & = & \bs{\Gamma}\left(t\right)\\
\hat{\bs K}\left(t\right) & = & -\bs K\left(t\right)
\end{eqnarray*}

\subsubsection{Retarded and advanced components }

Within the wide-band approximation the retarded and advanced self-energies
are local in time $\bs{\Sigma}_{c}^{R/A}(t,t')\propto\delta\left(t-t'\right)$
and $\bs G_{c}^{R/A}\left(t,t'\right)$ fulfills the differential
equation 
\begin{eqnarray*}
\left[i\pd_{t}-\bs K\left(t\right)\right]\bs G_{c}^{R}\left(t,t'\right) & = & \delta\left(t-t'\right)
\end{eqnarray*}
with boundary conditions 
\begin{eqnarray*}
\bs G_{c}^{R}\left(t,t'\right) & = & 0\text{ for }t'>t
\end{eqnarray*}
Solving the differential equation  gives 
\begin{eqnarray*}
\bs G_{c}^{R}\left(t,t'\right) & = & -i\Theta\left(t-t'\right)U\left(t,t'\right)
\end{eqnarray*}
and thus 
\begin{eqnarray*}
\bs G_{c}^{A}\left(t,t'\right) & = & i\Theta\left(t'-t\right)U^{\dagger}\left(t,t'\right)
\end{eqnarray*}
with 
\begin{eqnarray*}
\bs U\left(t,t'\right) & = & Te^{-i\int_{t'}^{t}d\tau\bs K\left(\tau\right)}\\
\bs U^{\dagger}\left(t,t'\right) & \equiv & \left[\bs U\left(t',t\right)\right]^{\dagger}=\bar{T}e^{i\int_{t'}^{t}d\tau\bs K^{\dagger}\left(\tau\right)}
\end{eqnarray*}
where $T$ and $\bar{T}$ are respectively the time-ordered and anti-time-ordered
operators.

\subsubsection{Keldysh component}

The Dyson Keldysh equations for the Keldysh component states that
\begin{eqnarray}
\left[\bs G_{c}^{R}\right]^{-1}\bs G_{c}^{K} & = & \bs{\Sigma}_{c}^{K}\bs G_{c}^{A}\label{eq:Kel_1}\\
\bs G_{c}^{K}\left[\bs G_{c}^{A}\right]^{-1} & = & \bs G_{c}^{R}\bs{\Sigma}_{c}^{K}\label{eq:Kel_2}
\end{eqnarray}
with $\left[\bs G_{c}^{R}\right]^{-1}\left(t,t'\right)=\delta\left(t,t'\right)\left[i\pd_{t}-\bs K\left(t\right)\right]$
and $\left[\bs G_{c}^{A}\right]\left(t,t'\right)=\left[\bs G_{c}^{R}\right]^{\dagger}\left(t',t\right)$.
In integral form we have 
\begin{eqnarray}
i\pd_{t}\bs G_{c}^{K}\left(t,t'\right) & = & \bs K\left(t\right)\bs G_{c}^{K}\left(t,t'\right)+i\int_{0}^{t}d\tau\,\bs{\Sigma}_{c}^{K}\left(t,\tau\right)U^{\dagger}\left(\tau,t'\right)\label{eq:G_K_diff_1}\\
-i\pd_{t'}\bs G_{c}^{K}\left(t,t'\right) & = & \bs G_{c}^{K}\left(t,t'\right)\bs K^{\dagger}\left(t'\right)-i\int_{0}^{t}d\tau\, U\left(t,\tau\right)\bs{\Sigma}_{c}^{K}\left(\tau,t'\right)\label{eq:G_K_diff_2}
\end{eqnarray}
The solution of these integral differential equations is given by
\begin{eqnarray}
\bs G_{c}^{K}\left(t,t'\right) & = & U\left(t,0\right)\bs G_{c}^{K}\left(0,0\right)U^{\dagger}\left(0,t'\right)+\int_{0}^{t}dt_{1}\int_{0}^{t'}dt_{2}U\left(t,t_{1}\right)\bs{\Sigma}_{c}^{K}\left(t_{1},t_{2}\right)U^{\dagger}\left(t_{2},t'\right)\label{eq:G_K_integrated}
\end{eqnarray}
where $\bs G_{c}^{K}\left(0,0\right)$ is the initial condition that
depends on the initial density matrix of the system.

\section{Quadratic Lindblad operators}

\subsubsection{Adjoint Lindblad equation }

The Lindblad equation for the evolution of the density matrix is given
by 

\begin{eqnarray}
\pd_{\tau}\rho\left(t\right) & = & \mathcal{L}\left[\rho\left(t\right)\right]\label{eq:H-1}\\
\mathcal{L}\left[\rho\right] & = & \mathcal{L}_{0}\left[\rho\right]+\sum_{\mu\neq0}\mathcal{L}_{\mu}\left[\rho\right]\\
\mathcal{L}_{0}\left[\rho\right] & = & -i\left[H,\rho\right]\\
\mathcal{L}_{\mu}\left[\rho\right] & = & \gamma_{\mu}\left(2L_{\mu}\rho L_{\mu}^{\dagger}-\left\{ L_{\mu}^{\dagger}L_{\mu},\rho\right\} \right)
\end{eqnarray}
where $H$ is the Hamiltonian of the system and $L_{\mu}$'s are due
to the interaction with the environment. 

Given the mean value of an observable $\mathcal{O}\left(t\right)=\tr\left[O\rho\left(t\right)\right]$
we define $\mathcal{L}^{\text{ad}}$ such that 
\begin{eqnarray}
\pd_{t}\mathcal{O}\left(t\right) & = & \tr\left\{ O\mathcal{L}\left[\rho\left(t\right)\right]\right\} \\
 & = & \tr\left\{ \mathcal{L}^{\text{ad}}\left[O\right]\rho\left(t\right)\right\} 
\end{eqnarray}
were we find by invariance of the trace 
\begin{eqnarray}
\mathcal{L}^{\text{ad}}\left[O\right] & = & \mathcal{L}_{0}^{\text{ad}}\left[O\right]+\sum_{\mu\neq0}\mathcal{L}_{\mu}^{\text{ad}}\left[O\right]\\
\mathcal{L}_{0}^{\text{ad}}\left[O\right] & = & i\left[H,O\right]\\
\mathcal{L}_{\mu}^{\text{ad}}\left[O\right] & = & \gamma_{\mu}\left\{ L_{\mu}^{\dagger}\left[O,L_{\mu}\right]+\left[L_{\mu}^{\dagger},O\right]L_{\mu}\right\} 
\end{eqnarray}
This means that the mean values of operators can be computed also
in the adjoint representation with 
\begin{eqnarray}
\mathcal{O}\left(t\right) & = & \tr\left[O\rho\left(t\right)\right]=\tr\left[O\left(t\right)\rho(0)\right]\\
\pd_{t}O\left(t\right) & = & \mathcal{L}^{\text{ad}}\left[O\left(t\right)\right]
\end{eqnarray}
which is the analog of the Heisenberg representation in usual Hamiltonian
dynamics.

\subsubsection{Lindblad equation for $\protect\bs{\chi}\left(t\right)$ }

For $H=\frac{1}{2}\bs C^{\dagger}.\bs H_{c}.\bs C$ and $L_{\ell}\left(t\right)=\sum_{i}\braket{\ell\left(t\right)}iC_{i}$,
one obtains, using the fermionic commutation relations, 
\begin{eqnarray}
\mathcal{L}_{0}^{\text{ad}}\left[\bs C\bs C^{\dagger}\right] & = & -i\left(\bs H_{c}\bs C\bs C^{\dagger}-\bs C\bs C^{\dagger}\bs H_{c}\right)\\
\sum_{\mu\neq0}\mathcal{L}_{\mu}^{\text{ad}}\left[\bs C\bs C^{\dagger}\right] & = & -\bs M\bs C\bs C^{\dagger}-\bs C\bs C^{\dagger}\bs M+\bs N
\end{eqnarray}
with 
\begin{eqnarray}
\bs M & = & \frac{1}{2}\left(\bs N+\bs J.\bs N^{*}.\bs J\right),\\
\bs N & = & \sum_{\mu}\ket{\ell_{\mu}}\gamma_{\mu}\bra{\ell_{\mu}}.
\end{eqnarray}
With the above expressions and for time dependent $\bs H_{c}$ and
$\bs N$ the evolution of the one body-density matrix: $\pd_{t}\bs{\chi}\left(t\right)=\tr\left\{ \mathcal{L}^{\text{ad}}\left[\left(\bs C\bs C^{\dagger}\right)(t)\right]\rho_{0}\right\} $,
can be written as 

\begin{eqnarray}
\pd_{t}\bs{\chi}\left(t\right) & = & -i\bs Q\left(t\right)\bs{\chi}\left(t\right)+i\bs{\chi}\left(t\right)\bs Q^{\dagger}\left(t\right)+\bs N\left(t\right)
\end{eqnarray}
with $\bs Q=\bs H_{c}-i\bs M$. This equation should be compared with
Eqs.(\ref{eq:G_K_diff_1}, \ref{eq:G_K_diff_2}). It can be integrated
similarly to Eq.(\ref{eq:G_K_integrated}) yielding to 
\begin{eqnarray}
\bs{\chi}\left(t\right) & = & U\left(t,0\right)\bs{\chi}\left(0\right)U^{\dagger}\left(0,t\right)+\int_{0}^{t}dt'U\left(t,t'\right)\bs N\left(t'\right)U^{\dagger}\left(t',t\right)\label{eq:Chi_of_t}
\end{eqnarray}

\subsubsection{Identification with the non-equilibrium Green's functions approach }

Setting $t'=t$ in Eq.(\ref{eq:G_K_integrated}) 
\begin{eqnarray}
i\pd_{t}\bs G_{c}^{K}\left(t,t\right) & = & \bs K\left(t\right)\bs G_{c}^{K}\left(t,t'\right)-\bs G_{c}^{K}\left(t,t\right)\bs K^{\dagger}\left(t\right)\label{eq:G_K_diff_1-1}\\
 &  & +i\int_{0}^{t}d\tau\,\left[\bs{\Sigma}_{c}^{K}\left(t,\tau\right)U^{\dagger}\left(\tau,t'\right)+U\left(t,\tau\right)\bs{\Sigma}_{c}^{K}\left(\tau,t'\right)\right]\nonumber 
\end{eqnarray}
and identifying $\bs{\chi}\left(t\right)$ by Eq.(\ref{eq:chi_GK})
we get
\begin{eqnarray}
\pd_{t}\bs{\chi}\left(t\right) & = & -i\bs K\left(t\right)\bs{\chi}\left(t\right)+i\bs{\chi}\left(t\right)\bs K^{\dagger}\left(t\right)+\frac{i}{2}\left[\bs K\left(t\right)-\bs K^{\dagger}\left(t\right)\right]\label{eq:G_K_diff_1-1-1}\\
 &  & +\frac{1}{2}i\int_{0}^{t}d\tau\,\left[\bs{\Sigma}_{c}^{K}\left(t,\tau\right)U^{\dagger}\left(\tau,t'\right)+U\left(t,\tau\right)\bs{\Sigma}_{c}^{K}\left(\tau,t'\right)\right]\nonumber 
\end{eqnarray}
and thus we may identify 
\begin{eqnarray*}
\bs Q\left(t\right) & = & \bs K\left(t\right)\\
\bs N\left(t\right) & = & \frac{i}{2}\left[\bs K\left(t\right)-\bs K^{\dagger}\left(t\right)\right]\\
 &  & +\frac{i}{2}\int_{0}^{t}d\tau\,\left[\bs{\Sigma}_{c}^{K}\left(t,\tau\right)U^{\dagger}\left(\tau,t\right)+U\left(t,\tau\right)\bs{\Sigma}_{c}^{K}\left(\tau,t\right)\right]
\end{eqnarray*}

\subsubsection{Wide-band regularization}

For the case of time independent quantities $\bs K\left(t\right)=\bs K$
and $\bs{\Gamma}_{\nu}\left(t,t'\right)=\bs{\Gamma}_{\nu}$ considered
in the main text the form of the operator $\bs N\left(t\right)=\sum_{\nu}\bs N_{\nu}\left(t\right)$
can be obtained explicitly: 
\begin{eqnarray}
\bs N_{\nu}\left(t\right) & = & \bs{\Gamma}_{\nu}+\hat{\bs{\Gamma}}_{\nu}\nonumber \\
 &  & +\left\{ \left[\int_{0}^{t}dt'e^{-i\bs K\left(t-t'\right)}F_{\nu}\left(t-t'\right)\right]\bs{\Gamma}_{\nu}-\left[\int_{0}^{t}dt'e^{-i\bs K\left(t-t'\right)}\bar{F}_{\nu}\left(t-t'\right)\right]\hat{\bs{\Gamma}}_{\nu}\right.\nonumber \\
 &  & +\left.\bs{\Gamma}_{\nu}\left[\int_{0}^{t}dt'F_{\nu}\left(t'-t\right)e^{i\bs K^{\dagger}\left(t-t'\right)}\right]-\hat{\bs{\Gamma}}_{\nu}\left[\int_{0}^{t}dt'e^{i\bs K^{\dagger}\left(t-t'\right)}\bar{F}_{\nu}\left(t'-t\right)\right]\right\} \label{eq:N-1}
\end{eqnarray}
To evaluate the integrals we use the regularization that amounts to
subtract the zero temperature result at a finite value of the reservoir
bandwidth $\Lambda$: 
\begin{eqnarray}
\int_{0}^{t}dt'e^{-i\bs K\left(t-t'\right)}F_{\nu}\left(t-t'\right) & = & \int_{0}^{t}dt'\int\frac{d\varepsilon}{2\pi}\left\{ \tanh\left[\beta_{\nu}\left(\varepsilon-\mu_{\nu}\right)\right]-\sgn\left[\varepsilon-\mu_{\nu}\right]\right\} e^{-i\left(\varepsilon+\bs K\right)t'}\nonumber \\
 &  & +\int_{0}^{t}dt'\int_{-\Lambda}^{\Lambda}\frac{d\varepsilon}{2\pi}\left\{ \sgn\left[\varepsilon\right]\right\} e^{-i\left(\varepsilon+\mu_{\nu}+\bs K\right)t'}
\end{eqnarray}
further simplifying we obtain 
\begin{eqnarray}
\int_{0}^{t}dt'e^{-i\bs K\left(t-t'\right)}F_{\nu}\left(t-t'\right) & = & iR\left[\left(\bs K+\mu_{\nu}\right),\beta_{\nu},t\right]-\frac{i}{\pi}\log\left(\Lambda t\right)
\end{eqnarray}
where 
\begin{eqnarray}
s\left[z,\tau\right] & = & -\int_{0}^{\tau}d\tau'\, e^{-iz\tau'}\int_{0}^{\infty}\frac{dx}{\pi}\left(\tanh\left[x\right]-1\right)\sin\left(x\tau'\right)\\
r\left[x\right] & = & \frac{\log\left[ix\right]+\Gamma\left[0,ix\right]+\gamma+\log\left(\frac{4}{\pi}\right)}{\pi}\\
R\left[\omega,\beta,t\right] & = & s\left[\beta\omega,t/\beta\right]+r_{0}\left[\omega t\right]
\end{eqnarray}
where $\Gamma\left[a,z\right]=\int_{z}^{\infty}dx\ x^{a-1}e^{-x}$
and $\gamma$ is the Euler constant. Note that in the expression for
$\bs N_{\nu}\left(t\right)$ the dependence on $\Lambda$ vanishes,
and the wide band limit is well defined, yielding to Eq.(\ref{eq:N_nu})
in the main text.

\subsubsection{Steady-state}

The equation for the steady state correlation matrix is given by 

\begin{eqnarray*}
0 & = & -i\bs K\bs{\chi}_{\infty}+i\bs{\chi}_{\infty}\bs K^{\dagger}+\bs N_{\infty}
\end{eqnarray*}
This equation can be solved explicitly considering the right and left
eigenvalues of $\bs K$ such that: 
\begin{eqnarray*}
\bs K & = & \sum_{\alpha}\ket{\alpha}\lambda_{\alpha}\bra{\alpha'}\\
\bs K^{\dagger} & = & \sum_{\alpha}\ket{\alpha'}\bar{\lambda}_{\alpha}\bra{\alpha}
\end{eqnarray*}
with the properties 
\begin{eqnarray*}
\sum_{\alpha}\ket{\alpha}\bra{\alpha'} & = & 1\\
\braket{\alpha}{\beta'} & = & \delta_{\alpha\beta}
\end{eqnarray*}
Inserting the partition of the identity into the equation for $\bs{\chi}_{\infty}$
we obtain 
\begin{eqnarray*}
\bra{\beta'}\bs{\chi}_{\infty}\ket{\gamma'} & = & -i\frac{\bra{\beta'}\bs N_{\infty}\ket{\gamma'}}{\lambda_{\beta}-\bar{\lambda}_{\gamma}}
\end{eqnarray*}
i.e. 
\begin{eqnarray*}
\bs{\chi}_{\infty} & =-i & \sum_{\beta\gamma}\ket{\beta}\frac{\bra{\beta'}\bs N_{\infty}\ket{\gamma'}}{\lambda_{\beta}-\bar{\lambda}_{\gamma}}\bra{\gamma}
\end{eqnarray*}

\section{Some details of Example II}

\subsection{Jordan-Wigner Transformed Hamiltonian }

Under a Jordan-Wigner transformation $\sigma_{m}^{+}=e^{i\pi\sum_{j=-\infty}^{m-1}c_{j}^{\dagger}c_{j}}c_{m}^{\dagger}$
the XY Hamiltonian becomes 

\begin{eqnarray*}
H & = & -\sum_{m}\frac{J}{2}\left[\left(1+\gamma\right)\sigma_{m}^{x}\sigma_{m+1}^{x}+\left(1-\gamma\right)\sigma_{m}^{y}\sigma_{m+1}^{y}\right]-h\sum_{m}\sigma_{m}^{z}\\
 & = & -\sum_{m}\frac{J}{2}\left[\left(2\gamma c_{m}^{\dagger}c_{m+1}^{\dagger}+2c_{m+1}^{\dagger}c_{m}+2c_{m}^{\dagger}c_{m+1}-2\gamma c_{m}c_{m+1}\right)\right]-h\sum_{m}\left(2c_{m}^{\dagger}c_{m}-1\right)
\end{eqnarray*}

\subsection{Observables}

For two observables $O_{i}=\frac{1}{2}\bs C^{\dagger}.\bs O_{i}.\bs C$
we have that 

\begin{eqnarray*}
\av{e^{z_{1}O_{1}}e^{z_{2}O_{2}}}_{t} & = & \sqrt{\det\left\{ \bs{\chi}\left(t\right)+e^{z_{1}\bs O_{1}}e^{z_{2}\bs O_{2}}\left[1-\bs{\chi}\left(t\right)\right]\right\} }
\end{eqnarray*}
varying with respect to $z_{1}$ and $z_{2}$
\begin{eqnarray*}
C_{O_{1},O_{2}} & = & \av{O_{1}O_{2}}_{t}-\av{O_{1}}_{t}\av{O_{2}}_{t}\\
 & = & \left.\pd_{z_{1}}\pd_{z_{2}}\ln\av{e^{z_{1}O_{1}}e^{z_{2}O_{2}}}_{t}\right|_{z_{1},z_{2}=0}\\
 & = & \frac{1}{2}\tr\left\{ \bs O_{1}\bs{\chi}\left(t\right)\bs O_{2}\left[1-\bs{\chi}\left(t\right)\right]\right\} 
\end{eqnarray*}
For the connected correlators along the $z$ direction, we obtain
\begin{eqnarray*}
C_{l,m} & = & \av{\sigma_{l}^{z}\sigma_{m}^{z}}_{t}-\av{\sigma_{l}^{z}}_{t}\av{\sigma_{m}^{z}}_{t}\\
 & = & \frac{1}{2}\tr\left\{ \mathcal{\bs S}_{l}\bs{\chi}\left(t\right)\mathcal{\bs S}_{m}\left[1-\bs{\chi}\left(t\right)\right]\right\} 
\end{eqnarray*}
with
\begin{eqnarray*}
\mathcal{\bs S}_{m} & = & \ket m\bra m-\ket{\hat{m}}\bra{\hat{m}}.
\end{eqnarray*}

Fig.(\ref{fig:XY_correlators}) shows the behavior of $C_{l,m}$ for
different values of $h$ and $\Delta h$ used to obtain the phase
diagram of Fig.(\ref{fig:XY}) in the main text.

\begin{figure}
\begin{centering}
\includegraphics[width=1\columnwidth]{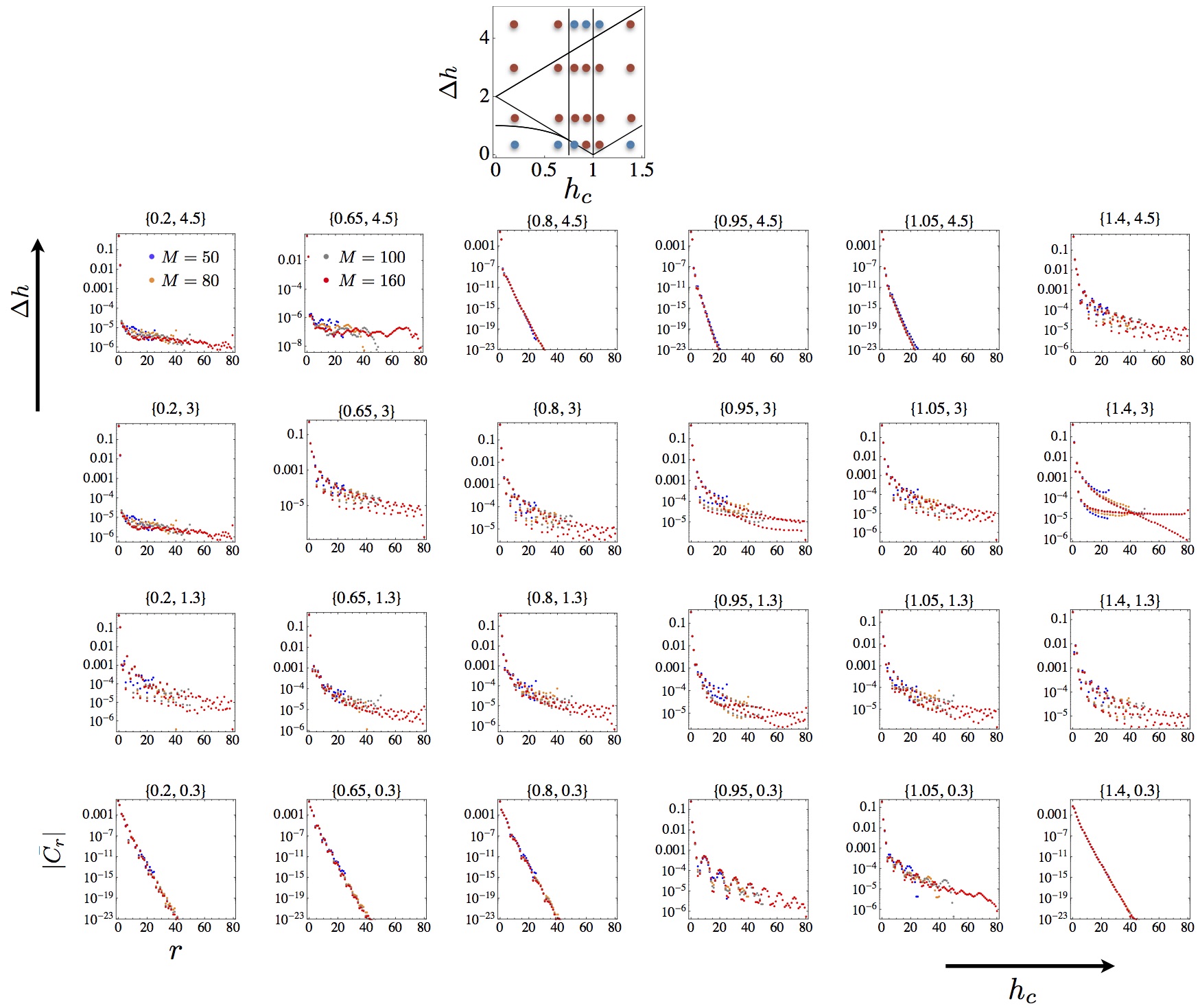}\protect\caption{\label{fig:XY_correlators}Upper panel: Non-equilibirum steady-state
phase diagram in the $h_{c}-\Delta h$ plane. The red (blue) dots
correspond to algebraic (exponential) spacial decay of the spin-spin
correlation function along the $z$ direction. Lower panel: Logarithmic
plots of the averaged correlation amplitudes $\bar{\protect\abs{C_{r}}}=\frac{2}{M}\sum_{m=M/2}^{M}\protect\abs{C_{r+m,m}}$
as a function of the distance between the spins, computed for different
values of $h$ and $\Delta h$. Note the clear distinction between
the algebraic and exponential decaying cases. }

\par\end{centering}

\end{figure}

\subsection{Currents}

Consider a partition $\Sigma$ of the complete system under analysis
with a finite range Hamiltonian and write the Hamiltonian as 
\begin{eqnarray*}
H & = & H_{\Sigma}+H_{\bar{\Sigma}}+H_{\pd\Sigma}
\end{eqnarray*}
where $H_{\Sigma}$ ($H_{\bar{\Sigma}}$ ) is the Hamiltonian restricted
to $\Sigma$ (the complement of $\Sigma$) and $H_{\pd\Sigma}$ collects
all the terms that are not separable in terms of $\Sigma$ and $\bar{\Sigma}$
degrees of freedom. A local quantity $Q$ is locally conserved if
the restriction of the observable $\hat{Q}$ to the region $\Sigma$
is conserved $\left[H_{\Sigma},\hat{Q}_{\Sigma}\right]=0$. The current
of the conserved quantity $Q$, leaving region $\Sigma$, is thus
given by 
\begin{eqnarray*}
\mathcal{J}_{Q,\Sigma} & = & -\frac{d}{dt}\av{Q_{\Sigma}}=-i\av{\left[H,\hat{Q}_{\Sigma}\right]}=-i\av{\left[H_{\pd\Sigma},\hat{Q}_{\Sigma}\right]}.
\end{eqnarray*}
Choosing $\Sigma$ to be a finite segment of an one dimensional system,
the boundary Hamiltonian is made of two disjoint pieces $H_{\pd\Sigma}=H_{\pd\Sigma_{L}}+H_{\pd\Sigma_{R}}$
corresponding to the left and right boundaries. The individual left
and right currents are thus given by 
\begin{eqnarray*}
\mathcal{J}_{Q,\Sigma}^{L/R} & = & -i\av{\left[H_{\pd\Sigma_{L/R}},Q_{\Sigma}\right]}.
\end{eqnarray*}
For steady-state conditions $\mathcal{J}_{Q,\Sigma}^{R}=-\mathcal{J}_{Q,\Sigma}^{L}$.

For the energy current of the XY model, with $\Sigma$ a segment of
the central region of Fig.\ref{fig:XY}-(a) , we have, in terms of
the Jordan-Wigner transformed fermions, 
\begin{eqnarray*}
Q_{\Sigma} & = & H_{\Sigma}=-\sum_{m:\left(m,m+1\in\Sigma\right)}\frac{J}{2}\left[\left(2\gamma c_{m}^{\dagger}c_{m+1}^{\dagger}+2c_{m+1}^{\dagger}c_{m}+2c_{m}^{\dagger}c_{m+1}-2\gamma c_{m}c_{m+1}\right)\right]-h\sum_{m\in\Sigma}\left(2c_{m}^{\dagger}c_{m}-1\right)\\
H_{\pd\Sigma_{R}} & = & -\frac{J}{2}\left[\left(2\gamma c_{m}^{\dagger}c_{m+1}^{\dagger}+2c_{m+1}^{\dagger}c_{m}+2c_{m}^{\dagger}c_{m+1}-2\gamma c_{m}c_{m+1}\right)\right]\text{ with }m\in\Sigma;\, m+1\in\bar{\Sigma}
\end{eqnarray*}
and 
\begin{eqnarray*}
\mathcal{J}_{Q,\Sigma}^{R} & = & \frac{1}{2}C^{\dagger}\left(\begin{array}{cc}
\bs j_{pp} & \bs j_{ph}\\
\bs j_{hp} & \bs j_{hh}
\end{array}\right)C
\end{eqnarray*}
with 
\begin{eqnarray*}
\bs j_{pp} & = & -iJ^{2}\left(1-\gamma^{2}\right)\left(\ket{m-1}\bra{m+1}-\ket{m+1}\bra{m-1}\right)-2ihJ\left(\ket{m-1}\bra m-\ket m\bra{m-1}\right)\\
\bs j_{hh} & = & iJ^{2}\left(1-\gamma^{2}\right)\left(\ket{\hat{m+1}}\bra{\hat{m-1}}-\ket{\hat{m-1}}\bra{\hat{m+1}}\right)+2ihJ\left(\ket{\hat{m}}\bra{\hat{m-1}}-\ket{\hat{m-1}}\bra{\hat{m}}\right)\\
\bs j_{hp} & = & 2i\gamma hJ\left(\ket{\hat{m}}\bra{m-1}-\ket{\hat{m-1}}\bra m\right)\\
\bs j_{ph} & = & -2i\gamma hJ\left(\ket{m-1}\bra{\hat{m}}-\ket m\bra{\hat{m-1}}\right)
\end{eqnarray*}
The mean value of the current operator can be computed as 
\begin{eqnarray*}
\av{\mathcal{J}_{Q,\Sigma}^{R}}_{t} & = & \frac{1}{2}\tr\left\{ \left(\begin{array}{cc}
\bs j_{pp} & \bs j_{ph}\\
\bs j_{hp} & \bs j_{hh}
\end{array}\right)\left[1-\bs{\chi}\left(t\right)\right]\right\} .
\end{eqnarray*}

\end{widetext}
\end{document}